\documentclass[11pt,twoside,a4]{article}
\usepackage{amssymb,amsfonts,amsbsy}
\newcommand{\be}{\begin{equation}}
\newcommand{\ee}{\end{equation}}
\newcommand{\ba}{\hspace*{-5pt}\begin{array}}
\newcommand{\ea}{\end{array}}
\newcommand{\p}{\partial}

\def\ord{\mathop{\rm ord}\nolimits}

\def\Im{\mathop{\rm Im}\nolimits}
\DeclareMathAlphabet{\bi}{OML}{cmm}{b}{it}

\def\diag{\mathop{\rm diag}\nolimits}

\def\Im{\mathop{\rm Im}\nolimits}

\newcommand{\bu}{\bi{u}}
\newcommand{\F}{\bi{F}}
\newcommand{\G}{\bi{G}}
\newcommand{\bG}{\bi{G}}

\newcommand{\bg}{\boldsymbol{\gamma}}
\def\ord{\mathop{\rm ord}\nolimits}
\def\Im{\mathop{\rm Im}\nolimits}

\def\diag{\mathop{\rm diag}\nolimits}
\def\tr{\mathop{\rm tr}\nolimits}

\def\res{\mathop{\rm res}\nolimits}

\newcommand{\ds}{\displaystyle}
\newtheorem{theo}{Theorem}
\newtheorem{lem}{Lemma}
\newtheorem{cor}{Corollary}

\begin{document}
         
\title
{\bf Locality of symmetries 
generated by nonhereditary, inhomogeneous,
and
time-de\-pen\-dent recursion~operators:\\
a new application for formal symmetries}
\author{{\protect\LARGE A. Sergyeyev}\\
Silesian University in Opava,
Mathematical Institute,\\ Bezru\v{c}ovo n\'am.~13,
746~01 Opava,
Czech Republic\\
E-mail: {\tt Artur.Sergyeyev@math.slu.cz}}
\date{}
\maketitle
\begin{abstract}
Using the methods of the theory of formal
symmetries, we obtain new easily verifiable
sufficient conditions for a recursion operator
to produce a hierarchy of local generalized
symmetries. An important advantage of our approach is that 
under certain mild assumptions it
allows to bypass the cumbersome check of hereditariness of the
recursion operator in question, what is
particularly useful for the study of symmetries of newly discovered
integrable systems. What is more, unlike the earlier work, 
the homogeneity of recursion operators and symmetries
under a scaling
is not assumed as well. An example of nonhereditary recursion operator
generating a hierarchy of local symmetries is presented. \looseness=-1
\end{abstract}

\section{Introduction}
Since the seminal paper of Olver \cite{olvjmp} it became clear
that the existence of a recursion operator  is one of the most
characteristic features of integrable systems of PDEs, cf.\
e.g.\ \cite{olv_eng2, bl, dor} and references therein.
%
Although the integrable hierarchies in (1+1) dimensions usually
prove to be local, this is not the case for their recursion
operators \cite{olv_eng2}--\cite{wang}.
Thus, it is natural
to ask under which conditions the repeated application of a
recursion operator to a local seed symmetry will yield {\em local}
symmetries corresponding to the higher equations of the
hierarchy.
\looseness=-1

To 
be more specific, consider first the simplest case of
one dependent variable $u$ and a
recursion operator of the form
\be\label{rosc}
\mathfrak{R}=\sum\limits_{i=0}^{r}a_{i}D^{i}+\sum
\limits_{\alpha=1}^{p} G_{\alpha} D^{-1}\circ \gamma_{\alpha},
\ee
where $r\geq 0$, $a_i$, $G_\alpha$, $\gamma_\alpha$
belong to the algebra 
$\mathcal{A}$ of locally analytic functions of $x,t,u,u_1,u_2,\dots$
(informally, $u_j=\p^j u/\p x^j$), 
and $D$ is the operator of total $x$-derivative, see below for details. 

It is immediate that because of the presence of $D^{-1}$ 
the application of $\mathfrak{R}$ to a $Q\in\mathcal{A}$
may yield a nonlocal expression, so the question is when the
repeated application of $\mathfrak{R}$ will not take us out of
$\mathcal{A}$. 
Clearly, proving that $\mathfrak{R}^j(Q)$ are 
local amounts to proving that
$\mathfrak{R}^j(Q)\gamma_\alpha \in\Im D$ for all $j$ and $\alpha$, 
where $\Im D$ stands for the image of $D$ in $\mathcal{A}$.
\looseness=-1
 
Adler~\cite{adl} proved this for the recursion
operators arising from zero-curvature representations of certain kind, 
and so did Olver \cite{olvbih} (cf.\ also Dorfman~\cite{dor})
for a class of recursion operators, associated with
bi-Hamiltonian systems. \looseness=-1

Moreover, in the bi-Hamiltonian case there is a very simple and
powerful proof of locality of $\mathfrak{R}^j(Q)$,
kindly communicated to the author by V.V. Sokolov and contained in his
unpublished work.
\looseness=-1
 
The idea 
of this proof 
is as follows: suppose 
that 
a recursion operator 
$\mathfrak{R}$ of the form (\ref{rosc}) with
$\p\mathfrak{R}/\p t= 0$ can be written 
as a ratio of two compatible time-independent 
Hamiltonian operators:
$\mathfrak{R}=\mathfrak{P}_1\mathfrak{P}_2^{-1}$.
Further assume that 
$\gamma_\alpha=\sum_{i=1}^{\infty}
c_{\alpha,i}\delta H_i/\delta u$
for some constants $c_{\alpha,i}$. Here $\delta/\delta u$ 
stands for the
variational derivative, and $H_i$ are
the Hamiltonians
recursively constructed via the Lenard scheme,  i.e., $\mathfrak{P}_1
\delta H_{i+1}/\delta u=\mathfrak{P}_2 \delta H_{i}/\delta u$,
cf.\ e.g.\ \cite{olv_eng2, dor}. 
\looseness-1

Under these assumptions, $H_i$ commute with respect
to the Poisson brackets associated with $\mathfrak{P}_1$ and
$\mathfrak{P}_2$. Hence if the seed symmetry $Q$
is Hamiltonian with respect to $\mathfrak{P}_1$ or 
$\mathfrak{P}_2$, and the respective Hamiltonian is a linear
combination of $H_i$, then $L_Q(\delta H_j/\delta u)=0$ for all $j$,
and hence $L_Q(\gamma_\alpha)=0$ 
and $\gamma_\alpha Q\in\Im D$, as
required.   Proceeding inductively, we can prove in this way that all
$Q_j\equiv\mathfrak{R}^j(Q)$  are local.
The same argument 
remains valid 
for the case of more than one dependent variable.
\looseness=-1

However, the above results leave open the question of how to prove the
locality of symmetries when the recursion operator 
does not fall into the realm of \cite{adl}
or we cannot (or, what is typical for the recursion operators of 
new integrable systems, we do not know how to) write it as a ratio of two
Hamiltonian operators. Recently, Sanders and Wang~\cite{swro} proved a
fairly general statement on locality of (1+1)-dimensional hierarchies of
evolution systems generated by hereditary recursion operators.
They prove that $\mathfrak{R}^j(Q)\gamma_\alpha \in\Im D$ using the
scaling-based arguments 
and the hereditariness of
$\mathfrak{R}$. However, their results
are limited to the case when both the recursion operators and 
symmetries they act on are 
homogeneous under certain scaling.
\looseness=-1


In the present paper we suggest a new approach which allows to overcome
this limitation and, unlike the earlier work, with minor changes can
be applied \cite{sergms} for the proof of 
locality of hierarchies generated
by master symmetries rather than recursion operators. Moreover,
under certain  mild assumptions our approach is applicable to
nonhereditary recursion operators as well, and, in particular, does not
require the check whether the operator in question is hereditary.  
\looseness=-1

We assume that
$\gamma_\alpha$ are linear combinations of variational derivatives of the
so-called {\em canonical conserved densities} \cite{mik1} 
constructed from 
$\mathfrak{R}$ (or, more broadly, from a nondegenerate formal symmetry
of infinite rank, see Theorem~\ref{th_loc} below). This is the case for
the majority of  known today recursion operators,
including e.g.\ those of KdV, mKdV,
nonlinear Schr\"odinger equation and many other integrable evolution
equations and systems. For instance, the recursion operator of KdV
equation
$$
\mathfrak{R}=D^2+(2/3)u D +(1/3) u_1 D^{-1}
$$
obviously is of the form (\ref{rosc}) with $p=1$ and
$\gamma_1=1$, the first nontrivial canonical density is 
$\rho_1=\res\mathfrak{R}^{1/2}=u/3$ and
$\gamma_1=1=(1/3)\delta\rho_1/\delta u$.
In fact, we failed to find an example that does not satisfy
our assumption within the class of recursion operators of the form 
(\ref{rosc}).
\looseness=-1

Let us stress that for a fairly large number of cases the
application of our results does not require the 
assumption that the recursion operator in
question is hereditary. Moreover, the verification of the conditions of
our theorems is considerably easier than the check of hereditariness,
because the latter usually is extremely cumbersome. This 
makes our results useful for the study of 
new recursion operators whose hereditariness is
not yet proved. We illustrate this in Section~5  by proving the locality
of a hierarchy generated by the recursion operator (\ref{bl_ro}).
\looseness=-1 


Below we present our main results in their simplest form, i.e.,
for the case of a single dependent variable $u$. In this case
the canonical densities are
$\rho_i=\res\mathfrak{R}^{i/r}$ for
$i=-1,1,2,\dots$, and
$\rho_0=(1/r)a_{r-1}/a_{r}$.
\looseness=-1 


\begin{theo}\label{th_wh_sc}
Let $\mathfrak{R}$ (\ref{rosc}) with $r>0$ 
be hereditary on the linear span of
$\mathfrak{R}^j(Q_0)$, $j=0,1,2\dots$, 
for a
$Q_0\in\mathcal{A}$ and satisfy $L_{Q_0}(\mathfrak{R})=0$,
and let there exist 
the functions $c_{\alpha,i}(t)$ such
that
\be\label{gammas_sc}\gamma_{\alpha}= \sum
_{i=-1}^{\infty}
c_{\alpha,i}(t)\delta\rho_{i}/\delta u
\ee for all
$\alpha=1,\dots,p$, where $\rho_i$ are canonical
densities associated with $\mathfrak{R}$.

Then $Q_{j}=\mathfrak{R}^{j}(Q_0)$  are local,
i.e., $Q_{j}\in\mathcal{A}$, for all $j\in\mathbb{N}$.
\end{theo}
\vspace{0pt}

We also can use the scaling-based arguments in order
to simultaneously 
prove the hereditariness of $\mathfrak{R}$ and the locality
of symmetries.
\begin{theo}\label{th_sc_scal}
Consider an operator
$\mathfrak{R}$ of the form (\ref{rosc}) with $r>0$, and let 
there exist the functions $c_{\alpha,i}(t)$ such that
(\ref{gammas_sc}) holds for all $\alpha=1,\dots,p$.
Let there also exist $Q_0\in\mathcal{A}$ and
$S\in\mathcal{A}$ such that $\ord Q_0\geq 2$, $\p Q_0/\p t=0$, 
$\p S/\p t=0$, $L_{Q_0}(\mathfrak{R})=0$,
$[S, Q_0]=\nu Q_0$, $\nu=\mathrm{const}$, $\nu\neq
0$, 
and $L_{S}(\mathfrak{R})=\zeta\mathfrak{R}$,
$\zeta=\mathrm{const}$, $\zeta\neq 0$.

Then
$Q_j=\mathfrak{R}^{j}(Q_0)$ are local, 
$L_{Q_j}(\mathfrak{R})=0$, 
$[Q_i,Q_j]=0$ for all $i,j=0,1,2\dots$, and $\mathfrak{R}$
is hereditary on the linear span of $Q_j$, $j=0,1,2,\dots$.
\looseness=-1
\end{theo}

Moreover, under certain conditions we can drop the requirement 
of hereditariness of $\mathfrak{R}$,
but then we have to fix an
evolution equation
\be\label{eveq_sc}
u_t=F(x,t,u,\dots, u_n), n\geq 2.
\ee 
$$
\hspace*{-52mm}\mbox{Let}\,\,n_0=\left\{\begin{array}{l} 1-j,\,
\mbox{where
$j$
is the greatest integer such that}\\
\quad\p F/\p u_i\,\,\mbox{for}\,\,
i=n-j,\dots,n\,\,\mbox{depend
on
$x$ and $t$ only},\\ 2\,\,\mbox{otherwise.} \end{array} \right.
$$  
\begin{theo}\label{th_loc_sc}
Let (\ref{eveq_sc}) with $n\geq 2$
have a 
symmetry $Q_0\in\mathcal{A}$ such that
$\ord Q_0\geq\max(n_0,0)$ and $\p Q_0/\p t=0$, and a 
recursion operator
$\mathfrak{R}$ of the form (\ref{rosc}) with $r>0$ such that
$\p\mathfrak{R}/\p t=0$. Further assume that (\ref{gammas_sc}) holds 
and $c_{\alpha,i}(t)\equiv 0$ for all $i>n-n_0-1$ and
all $\alpha=1,\dots,p$.

Then $Q_{j}=\mathfrak{R}^j(Q_0)$ are local
for all $j\in\mathbb{N}$.
\end{theo}

Note that if $\p F/\p t=0$, we can take $F$ for $Q_0$ in
Theorem~\ref{th_loc_sc}.

Let us also mention that if $\p F/\p t=0$, then the densities 
$\rho_i$ for
$i=-1,\dots,n-n_0-1$, unlike their counterparts 
with $i> n-n_0-1$, can be
computed directly from $F$ \cite{mik1,s}: 
$\rho_0=(1/n)\p F/\p u_{n-1}(\p
F/\p u_{n})^{-1}$,  and for $j=-1,1\dots,n-n_0-1$  
$\rho_j=\res\mathfrak{R}^{j/r}$ is a linear combination 
of $\tilde\rho_k$,
$k=-1,1,\dots,j$, where 
$\tilde\rho_k=\res (F')^{k/n}$,
$F'=\sum_{i=0}^{n}\p F/\p u_i D^i$.   
\looseness=-1


In the rest of the paper we present 
the extension of Theorems~\ref{th_wh_sc}--~\ref{th_loc_sc} 
to the case of more than one dependent variable, see Theorems
\ref{th_wh}--\ref{th_loc} below, along with the complete proofs and
examples, see Section 5 for the latter.
In this case the coefficients of recursion operators
are matrices,  so the nondegeneracy and diagonalizability issues come
into play. 
Nevertheless, the extra conditions that appear in our theorems
in the case of more than one dependent variable are easy to verify,
and they are satisfied for a large number of interesting examples. 
\looseness=-1

Our technique can be applied not only
to the recursion operators but also to the so-called {\em inverse
Noether operators} \cite{ff} that map symmetries to cosymmetries.
Namely, 
our Corollaries~\ref{cor_wh} and \ref{cor_loc} enable one to prove
the existence of infinite number of {\em local} cosymmetries, and
eventually of local conserved densities, 
generated using these operators.
Note that whenever exist, the
symplectic operators
of integrable evolution systems automatically are \cite{ff}
inverse Noether operators for these systems. 
\looseness=-1

\section{Preliminaries}
Consider an algebra $\mathcal{A}_{j}$ of locally analytic
functions of $x,t,\bu,\bu_{1},\dots,\bu_{j}$, where
$\bu_{k}=(u_{k}^{1},\dots,u_{k}^{s})^{T}$ are vectors,
$\bu_0\equiv\bu$, and let $\mathcal{A}=\cup
_{j=0}^{\infty}\mathcal{A}_{j}$. We shall call the elements of
$\mathcal{A}$ {\em local} functions, cf.\ e.g.\ \cite{olv_eng2,
sok}. Let us further
introduce \cite{olv_eng2, sok} a derivation
$
D\equiv D_x=\p/\p x+\sum
_{j=0}^{\infty}\bu_{j+1}\p/\p\bu_j$,
the
variational derivative
$\delta/\delta\bu=\sum
_{j=0}^{\infty}(-D)^{j}\p/\p\bu_j$ and the operator of directional
derivative
of $\vec f\in\mathcal{A}^{q}$:
$\vec f'=\sum
_{i=0}^{\infty}\p \vec f/\p\bu_{i}D^{i} $. In what follows we
shall see that $x$ will play the role of space variable, $D$ will
be the total $x$-derivative, and, if we specify an evolution system
$\p\bi{u}/\p t=\bi{F}(x,t,\bu,\dots,\bu_n)$, then
$t$ becomes an evolution parameter. \looseness=-1

The Lie bracket, see e.g.\ \cite{olv_eng2}, given for any
$\bi{P},\bi{Q}\in\mathcal{A}^s$ by
$[\bi{P},\bi{Q}]=\bi{P}'[\bi{Q}]-\bi{Q}'[\bi{P}]$ 
makes the linear space $\mathcal{A}^s$ of $s$-component vectors
whose components belong to $\mathcal{A}$ into a Lie algebra, 
and enables us to
define the Lie derivative of $\bi{R}\in\mathcal{A}^s$ along
$\bi{Q}\in\mathcal{A}^s$ by setting
$L_{\bi{Q}}(\bi{R})=[\bi{Q},\bi{R}]$.

Now consider \cite{olv_eng2, mik1, s} the set
$\mathrm{Mat}_{q}(\mathcal{A})[\![D^{-1}]\!]$ of formal series of the
form $\mathfrak{H}=\sum
_{j=-\infty}^{k}h_{j}D^{j}$, where $h_{j}$ are $q\times q$
matrices with entries from $\mathcal{A}$. We shall write
$\mathcal{A}[\![D^{-1}]\!]$ instead of
$\mathrm{Mat}_{1}(\mathcal{A})[\![D^{-1}]\!]$ for short. The
greatest $m$ such that $h_{m}\neq 0$ is called the {\em degree} of
formal series $\mathfrak{H}$ and is denoted as $\deg\mathfrak{H}$,
cf.\ e.g.\ \cite{olv_eng2, mik1, s}.
Clearly, $\G'\in
\mathrm{Mat}_{s}(\mathcal{A})[\![D^{-1}]\!]$ for any
$\G\in\mathcal{A}^{s}$, so we shall define the {\em order} of
$\G\in\mathcal{A}^s$  as $\ord\G=\deg\G'$.

A formal series $\mathfrak{H}=\sum
_{j=-\infty}^{m}h_{j}D^{j}
\in\mathrm{Mat}_{q}(\mathcal{A})[\![D^{-1}]\!]$ of degree $m$ is
called {\em non\-de\-ge\-nerate} \cite{mik1} if $\det h_{m}\neq
0$.
For $\mathfrak{H}=\sum
_{j=-\infty}^{m}h_{j}D^{j}
\in
\mathcal{A}[\![D^{-1}]\!]$, $\deg\mathfrak{H}=m$, define its
{\em residue} as $\res\mathfrak{H}=h_{-1}$, and
its {\em logarithmic residue} as $\res\ln\mathfrak{H}=h_{m-1}/h_{m}$,
see e.g.\ \cite{mik1}.
%
For $\mathfrak{H}=\sum
_{j=-\infty}^{m}h_{j}D^{j}
\in\mathrm{Mat}_{q}(\mathcal{A})[\![D^{-1}]\!]$
and $\bi{Q}\in\mathcal{A}^{s}$ set
$\mathfrak{H}'[\bi{Q}]=\sum
_{j=-\infty}^{m}h'_{j}[\bi{Q}]D^{j}$.

The multiplication law, 
which is nothing but the generalized Leibniz rule,
cf.\ e.g.\ \cite{olv_eng2},
\be\label{lei} a D^{i}\circ b D^{j} =a
\sum\limits_{q=0}^{\infty}{\displaystyle \frac{i(i-1)\cdots
(i-q+1)}{q!}}D^{q}(b)D^{i+j-q}, \ee
extended by linearity to the whole
$\mathrm{Mat}_{q}(\mathcal{A})[\![D^{-1}]\!]$, makes this set into
an algebra, and the commutator $\left[ \mathfrak{A}, \mathfrak{B}
\right]=\mathfrak{A} \circ \mathfrak{B}- \mathfrak{B} \circ
\mathfrak{A}$ further makes it into a Lie algebra. 
Below we omit $\circ$ unless this leads to confusion.

For any nondegenerate
$\mathfrak{H}\in\mathrm{Mat}_{q}(\mathcal{A})[\![D^{-1}]\!]$ we
can \cite{mik1} define its inverse
$\mathfrak{H}^{-1}\in\mathrm{Mat}_{q}(\mathcal{A})[\![D^{-1}]\!]$
such that
$\mathfrak{H}\circ\mathfrak{H}^{-1}=
\mathfrak{H}^{-1}\circ\mathfrak{H}=\mathbb{I}$,
where $\mathbb{I}$ is a $q\times q$ unit matrix. Moreover
\cite{olv_eng2, mik1, sok, s}, for any
$\mathfrak{H}\in\mathcal{A}[\![D^{-1}]\!]$,
$\deg\mathfrak{H}=m\neq 0$, we can define
its $m$th root $\mathfrak{H}^{1/m}$ and its fractional powers
$\mathfrak{H}^{j/m}=(\mathfrak{H}^{1/m})^{j}$, and these
fractional powers commute:
$[\mathfrak{H}^{i/m},\mathfrak{H}^{j/m}]=0$ for all
$i,j\in\mathbb{Z}$ \cite{olv_eng2, mik1}.

Define \cite{olv_eng2} the Lie derivative of 
$\mathfrak{H}\in\mathrm{Mat}_{s}(\mathcal{A})[\![D^{-1}]\!]$ along
$\bi{Q}\in\mathcal{A}^{s}$ as 
$L_{\bi{Q}}(\mathfrak{H})= \mathfrak{H}'[\bi{Q}] 
- [\bi{Q}',\mathfrak{H}]$.
A formal series
$\mathfrak{H}\in\mathrm{Mat}_{s}(\mathcal{A})[\![D^{-1}]\!]$ is
called \cite{olv_eng2, mik1, sok,s}  a
{\em formal symmetry} of rank $m$ (respectively of infinite rank)
for an evolution system $\p\bu/\p\tau=\bi{Q}$,
$\bi{Q}\in\mathcal{A}^{s}$, where $\tau$ is an evolution parameter
other than $t$, if $\deg L_{\bi{Q}}(\mathfrak{H})\leq
\deg\mathfrak{H}+\deg\bi{Q}'-m$ (resp.\
$L_{\bi{Q}}(\mathfrak{H})=0$).
The formal symmetry for an evolution system $\p\bu/\p t=\bi{Q}$
is defined in a similar way:
we only should \cite{mik1, sok, s}
replace $L_{\bi{Q}}(\mathfrak{H})$ by
$\p\mathfrak{H}/\p t+ L_{\bi{Q}}(\mathfrak{H})$
in the above definition.
%
%
%

\begin{lem}\label{lem1}
Consider 
$\mathfrak{H}=\sum
_{j=-\infty}^{m}h_{j}D^{j}
\in\mathrm{Mat}_{s}(\mathcal{A})[\![D^{-1}]\!]$
such that $\deg\mathfrak{H}=m\neq\nobreak 0$ and the matrix $h_m$ has
exactly $s$ distinct eigenvalues $\lambda_i$. Let $\Gamma$ be a
matrix bringing $h_m$ into the diagonal form:
$h_{m}=\Gamma^{-1}\Lambda\Gamma$,
$\Lambda=\diag(\lambda_1,\dots,\lambda_s)$.

Then there exists a unique formal series
$
\mathfrak{T}=\Gamma+ \sum
_{j=-1}^{-\infty}\Gamma_{j}\Gamma D^{j}$
with the property $\diag\Gamma_{j}=0$,
$j=-1,-2,\dots$, such that
all coefficients of the formal series
$\tilde\mathfrak{H}=\mathfrak{T}\mathfrak{H}\mathfrak{T}^{-1}$ are
diagonal matrices; $\Gamma_j$ are $s\times s$ matrices with
entries from $\mathcal{A}$. 

Moreover, for any
$\bi{Q}\in\mathcal{A}^{s}$, 
$\ord\bi{Q}\geq 1$, 
such that
$\mathfrak{H}$ is the formal symmetry of infinite rank for the
equation $\p\bu/\p\tau=\bi{Q}$ (resp.\ $\p\bu/\p t=\bi{Q}$), all
coefficients of the formal series $\mathfrak{V}\equiv
\mathfrak{T}\bi{Q}'\mathfrak{T}^{-1} + \mathfrak{T}'[\bi{Q}]
{\mathfrak{T}}^{-1}$ (resp.\ $\mathfrak{W}\equiv
\mathfrak{T}\bi{Q}'\mathfrak{T}^{-1} + (\p\mathfrak{T}/\p t
+\mathfrak{T}'[\bi{Q}]){\mathfrak{T}}^{-1}$) are diagonal
matrices.\looseness=-1
\end{lem}
\noindent{\it Proof.}
Equating the coefficients at the powers of $D$ in
$\mathfrak{T}\mathfrak{H}=\tilde\mathfrak{H}\mathfrak{T}$ enables
us to find $\Gamma_j$ and the coefficients of
$\tilde\mathfrak{H}$, whence we readily infer the existence and
uniqueness of $\mathfrak{T}=\Gamma+\sum
_{j=-1}^{-\infty}\Gamma_{j}\Gamma D^{j}$ such that
$\diag\Gamma_{j}=0$ and all coefficients of
$\tilde\mathfrak{H}=\mathfrak{T}\mathfrak{H}\mathfrak{T}^{-1}$ are
diagonal matrices,
cf.\ the proof of Proposition~2.1 in \cite{mik1}. We use here the
well-known fact that for any diagonal $s\times s$ matrix $\Lambda$
with $s$ distinct eigenvalues and for any $s\times s$ matrix
$\Omega$ with zero diagonal entries (we write this as $\diag\Omega=0$)
the
equation $[\Lambda,\Psi]=\Omega$ has a unique solution in the class of
$s\times s$ matrices $\Psi$ with $\diag\Psi=0$,
cf.\ e.g.\
\cite{mik1}.
\looseness=-1

Further observe that the transformation
$\mathfrak{H}\rightarrow\tilde\mathfrak{H}$ and
$\bi{Q}'\rightarrow \mathfrak{V}$
(resp.\ $\bi{Q}'\rightarrow \mathfrak{W}$
for the equation $\bu_t=\bi{Q}$) takes the equation
$L_{\bi{Q}}(\mathfrak{H})=0$ (resp.\ $\p\mathfrak{H}/\p
t+L_{\bi{Q}}(\mathfrak{H})=0$) to
$\tilde\mathfrak{H}'[\bi{Q}]-[\mathfrak{V},\tilde\mathfrak{H}]=0$
(resp.\ to  $\p\tilde\mathfrak{H}/\p t+
\tilde\mathfrak{H}'[\bi{Q}]-[\mathfrak{W},\tilde\mathfrak{H}]=0$).
%
Using the above-mentioned result on solutions of
$[\Lambda,\Psi]=\Omega$ and successively solving the equations for
the coefficients of $\mathfrak{V}$ (resp.\ $\mathfrak{W}$) that arise
from equating to zero the coefficients at powers of $D$ on the left-hand
side of the equation
$\tilde\mathfrak{H}'[\bi{Q}]-[\mathfrak{V},\tilde\mathfrak{H}]=0$
(resp.\  $\p\tilde\mathfrak{H}/\p t+
\tilde\mathfrak{H}'[\bi{Q}]-[\mathfrak{W},\tilde\mathfrak{H}]=0$),
we readily see that all coefficients of $\mathfrak{V}$ (resp.\
$\mathfrak{W}$)
are diagonal matrices. $\square$
\looseness=-1
Following \cite{mik1}, consider the {\em canonical densities}
associated with $\mathfrak{H}$:
$\rho_{j}^{a}=
\allowbreak\res((\mathfrak{T}\mathfrak{H}
\mathfrak{T}^{-1})_{aa})^{j/|m|}$,
$j=-1,1,2,\dots$, $a=1,\dots,s$, and $\rho_{0}^{a}=\res\ln
((\mathfrak{T}\mathfrak{H}\mathfrak{T}^{-1})_{aa})^{1/|m|}$,
$a=1,\dots,s$, where the subscript `$aa$' means taking the
$(a,a)$-th entry of the matrix. Lemma 1 and the results of
\cite{mik1} readily imply the following assertion.
\looseness=-1
\begin{cor}\label{cons}
Under the assumptions of Lemma 1, suppose that $\mathfrak{H}$ is
nondegenerate. Then the quantities $\rho_{j}^{a}$, $a=1,\dots,s$,
$j=-1,0,1,2,\dots$ are conserved densities for the equation
$\bu_{\tau}=\bi{Q}$ (resp.\ $\bu_{t}=\bi{Q}$), i.e.,
${\rho'}_{j}^{a}[\bi{Q}]\in\Im D$ (resp.\  $\p\rho_{j}^a/\p t +
{\rho'}_{j}^a[\bi{Q}]\in\Im D$).\looseness=-1
\end{cor}
\looseness=-1

Thus, the canonical densities associated with $\mathfrak{H}$
are common conserved densities for all systems of the form
$\bi{u}_{\tau}=\bi{Q}$ (or $\bi{u}_t=\bi{Q}$) admitting 
$\mathfrak{H}$ as
a formal symmetry of infinite rank.  Moreover, the formal series
$\mathfrak{T}$ that diagonalizes $\mathfrak{H}$ simultaneously
diagonalizes the formal series $\mathfrak{V}$ 
(or $\mathfrak{W}$) for all
these systems.
\looseness=-1
\section{The main result}
Consider an operator
\be\label{ro}
\mathfrak{R}=\sum\limits_{i=0}^{r}a_{i}D^{i}+\sum
\limits_{\alpha=1}^{p}\G_{\alpha}\otimes D^{-1}\circ \bg_{\alpha},
\ee
where $r\geq 0$, $a_i$ are $s\times s$ matrices with entries from
$\mathcal{A}$, and $\bG_{\alpha},\bg_{\alpha}\in\mathcal{A}^{s}$,
i.e., $\bG_{\alpha},\bg_{\alpha}$ are $s$-component vectors with
entries from $\mathcal{A}$.

The Lie derivative $L_{\bi{Q}}(\mathfrak{R})$ of $\mathfrak{R}$
along $\bi{Q}\in\mathcal{A}^s$ is given \cite{olv_eng2} by
$L_{\bi{Q}}(\mathfrak{R})=\mathfrak{R}'[\bi{Q}] -
[\bi{Q}',\mathfrak{R}]$, that is, by the very same formulae as for the
elements of $\mathrm{Mat}_{s}(\mathcal{A})[\![D^{-1}]\!]$,
because using the generalized
Leibniz rule (\ref{lei})
we always can rewrite $\mathfrak{R}$ as infinite formal series in
powers of $D$, and thus consider it as an element of
$\mathrm{Mat}_{s}(\mathcal{A})[\![D^{-1}]\!]$.
\looseness=-1

An operator $\mathfrak{R}$ is called, see e.g.
\cite{olv_eng2}, a {\em recursion operator} for an evolution
system $\bi{u}_{t}=\bi{F}(x,t,\bi{u},\dots,\allowbreak\bi{u}_n)$
(respectively
$\bi{u}_{\tau}=\bi{F}(x,t,\allowbreak\bi{u},\dots,\bi{u}_n)$, where
$\tau$ is an evolution parameter other than $t$), if it is a formal
symmetry of infinite rank for this system, i.e. if it satisfies the
equation
$\p\mathfrak{R}/\p t +
L_{\bi{F}}(\mathfrak{R})=0$ (resp.\ $L_{\bi{F}}(\mathfrak{R})=0$).
\looseness=-1

Note that nearly all known today recursion operators of integrable
(1+1)-di\-men\-si\-on\-al evolution systems are 
\cite{swro, wang} of the form
(\ref{ro}), with only a few  
exceptions, see e.g.\
\cite{gut,ma,sak}.
\looseness=-1

Recall that a linear operator $\mathfrak{R}$ is said \cite{ff} to
be {\it hereditary} (or {\it Nijenhuis} \cite{dor}) on a linear
space $\mathcal{L}$,
if for all $\bi{Q}\in\mathcal{L}$ 
\be\label{heredcond}
L_{\mathfrak{R}(\bi{Q})}(\mathfrak{R})=\mathfrak{R}\circ
L_{\bi{Q}}(\mathfrak{R}). 
\ee 

In what follows by saying that a recursion operator is hereditary
without specifying $\mathcal{L}$ we shall mean that it is
hereditary on its whole domain of definition, cf.\ \cite{swro}. 
If $\mathfrak{R}$ is hereditary on $\mathcal{L}$, 
then \cite{ff} for any $\bi{Q}\in\mathcal{L}$ we have
$[\mathfrak{R}^{i}(\bi{Q}),\mathfrak{R}^{j}(\bi{Q})]=0$,
$i,j=0,1,2,\dots$.

Let $\mathcal{S}(\mathfrak{R},\bi{Q})$ denote the linear span of
$\mathfrak{R}^i(\bi{Q})$, $i=0,1,2,\dots$. It is immediate from
(\ref{heredcond}) that
$L_{\mathfrak{R}^i(\bi{Q})}(\mathfrak{R})=0$ for all
$i=0,1,2,\dots$
if and only if $L_{\bi{Q}}(\mathfrak{R})=0$ and
$\mathfrak{R}$ is hereditary on
$\mathcal{S}(\mathfrak{R},\bi{Q})$. In view of this, we have
$[\mathfrak{R}^{i}(\bi{Q}),\mathfrak{R}^{j}(\bi{Q})]=0$ for all
$i,j=0,1,2,\dots$, provided \be\label{wh}
L_{\mathfrak{R}^i(\bi{Q})}(\mathfrak{R})=0\,\,\, \mbox{for
all}\,\,\, i=0,1,2,\dots. \ee

In view of this, the results of Sanders and Wang \cite{swro} on
locality of hierarchies of symmetries generated by recursion operators
obviously remain valid if the requirement of
hereditariness of recursion operator on its whole domain of
definition
is replaced by (\ref{wh}) with $i=1,2,\dots$. 
What is more,
(\ref{wh}) can be often proved using 
simple scaling arguments (which are,
however, quite different from those of \cite{swro}), see
Theorem~\ref{th_sc} below. \looseness=-1

Let us now turn to a general situation when (\ref{wh})
holds, but, unlike~\cite{swro},
no assumption on scaling properties of $\mathfrak{R}$ and $\bi{Q}$
is made.

\begin{theo}\label{th_wh}
Let $\mathfrak{R}$ (\ref{ro}) with $r>0$ 
be hereditary on $\mathcal{S}(\mathfrak{R},\bi{Q}_0)$ for a
$\bi{Q}_0\in\mathcal{A}^{s}$. 
Suppose that $L_{\bi{Q}_0}(\mathfrak{R})=0$,
$\det a_r\neq 0$, the matrix $a_r$ has $s$ distinct
eigenvalues, and 
there exist the functions $c_{\alpha,i,a}(t)$ such
that
\be\label{gammas}\bg_{\alpha}= \sum
_{i=-1}^{\infty}\sum
_{a=1}^{s}c_{\alpha,i,a}(t)\delta\rho_{i}^{a}/\delta\bu
\ee for all
$\alpha=1,\dots,p$, where $\rho_i^a$ are canonical
densities associated with $\mathfrak{R}$.
\looseness=-1

Then $\bi{Q}_{j}=\mathfrak{R}^{j}(\bi{Q}_0)$  are local,
that is, $\bi{Q}_{j}\in\mathcal{A}^{s}$, for all
$j\in\mathbb{N}$, and commute: $[\bi{Q}_i,\bi{Q}_j]=0$ for all
$i,j=0,1,2\dots$.
\end{theo}
\noindent{\it Proof.} As $\mathfrak{R}$ is hereditary on
$\mathcal{S}(\mathfrak{R},\bi{Q}_0)$, we have
$L_{\mathfrak{R}^j(\bi{Q}_0)}(\mathfrak{R})=0$, and thus 
$\mathfrak{R}$ is a formal symmetry 
of infinite rank for all evolution
systems of the form
$\p\bu/\p t_{j}=\bi{Q}_{j}$,
$j=0,1,2,\dots$. Hence, if $r>0$, then the canonical densities
$\rho_i^a$ for
$i=-1,0,1,2,\dots$ and $a=1,\dots,s$ 
are conserved densities for all these systems.
%
If $\bi{Q}_j$ is local, this means that 
${\rho'}_{i}^{a}[\bi{Q}_j]\in\Im D$. But 
${\rho'}_{i}^{a}[\bi{Q}_j]\in\Im D$
if and only if
$\delta\rho_{i}^{a}/\delta\bu\cdot\bi{Q}_{j}\in\Im D$,
where $\cdot$ stands for the scalar product of $s$-component
vectors.
\looseness=-1

Hence, $\bg_{\alpha}\cdot\bi{Q}_j\in\Im D$ by (\ref{gammas}), 
because
$\delta\rho_{i}^a/\delta\bu\cdot\bi{Q}_j\in\Im D$ by the above.
Thus, $D^{-1} (\bg_{\alpha}\cdot{\bi{Q}}_j)\in\mathcal{A}$ for all
$\alpha=1,\dots,p$, and $\bi{Q}_{j+1}$ is local, if so is $\bi{Q}_j$.
The induction on $j$ starting from $j=0$
completes the proof. $\square$ \looseness=-1

\noindent{\it Remark 1.} The condition $\det a_r\neq 0$ can be relaxed.
Indeed, if $\det a_r=0$, then the matrix $a_r$ has precisely one
zero eigenvalue, say $\lambda_b$: $\lambda_b=0$. Then we should
just require that $ m\equiv
\deg(\mathfrak{T}\mathfrak{R}\mathfrak{T}^{-1})_{bb}\neq 0$ and
replace the densities $\rho_{i}^{b}$ by the following ones:
$\tilde\rho_{j}^{b}=
\res((\mathfrak{T}\mathfrak{R}\mathfrak{T}^{-1})_{bb})^{j/|m|}$
for $j=-1,1,2,\dots$, and $\tilde\rho_{0}^{b}=\res\ln
((\mathfrak{T}\mathfrak{R}\mathfrak{T}^{-1})_{bb})^{1/|m|}$.

\noindent{\it Remark 2.} If we have $\bg_{\alpha}= \sum
_{i=-1}^{\infty}c_{\alpha,i}(t)\delta\tilde\rho_{i}/\delta\bu$ for
all $\alpha=1,\dots,p$, where
$\tilde\rho_i=\tr\res\mathfrak{R}^{i}$, $i=-1,1,2,\dots$,
$\tilde\rho_0=\tr\res\ln\mathfrak{R}$, then the assumption of
Theorem 1 that $a_r$ has $s$ distinct eigenvalues can be dropped.
Furthermore, if $c_{\alpha,-1}(t)=c_{\alpha,0}(t)=0$ for all
$\alpha=1,\dots,p$,
then we can also drop the assumption $\det a_r\neq 0$.

As we have already mentioned above, it is possible to 
prove that $\mathfrak{R}$
is hereditary on $\mathcal{S}(\mathfrak{R},\bi{Q}_0)$ using the
scaling-based arguments. The locality of
$\bi{Q}_i$ then follows by the same argument as in the
proof of Theorem~\ref{th_wh}. 
\looseness=-1
\begin{theo}\label{th_sc}
Let $\mathfrak{R}$ of the
form (\ref{ro}) with $r>0$ be such that
$\p\mathfrak{R}/\p t=0$, $\det a_r\neq 0$,
the matrix $a_r$ has $s$ distinct eigenvalues,
and let there exist the functions $c_{\alpha,i,a}(t)$ such that 
(\ref{gammas}) holds 
for all
$\alpha=1,\dots,p$.
Let there also exist  $\bi{Q}_0\in\mathcal{A}^s$
and  $\bi{S}\in\mathcal{A}^s$
such that 
$q_0\equiv\ord\bi{Q}_0\geq 2$, 
$\p \bi{Q}_0/\p t=0$, 
the matrix $\p\bi{Q}_0/\p\bi{u}_{q_0}$ is nondegenerate and
has $s$ distinct eigenvalues, 
$\p\bi{S}/\p t=0$,
$L_{\bi{Q}_0}(\mathfrak{R})=0$,
$[\bi{S},\bi{Q}_0]=\nu\bi{Q}_0$,
where 
$\nu$ is a nonzero constant.
Further assume that 
$L_{\bi{S}}(\mathfrak{T}^{-1}c\tilde\mathfrak{R}^{1/r}\mathfrak{T})=
\mu\mathfrak{T}^{-1}c\tilde\mathfrak{R}^{1/r}\mathfrak{T}$
for any matrix $c=\diag(c_1,\dots,c_s)$, $c_i={\rm const}$, where
$\mu$ is a nonzero constant independent of $c_1,\dots,c_s$,  
$\tilde\mathfrak{R}=\mathfrak{T}\mathfrak{R}\mathfrak{T}^{-1}$,
and $\mathfrak{T}$ is the formal series constructed in Lemma
\ref{lem1} for $\mathfrak{H}=\mathfrak{R}$.
    
Then
$\bi{Q}_j=\mathfrak{R}^{j}(\bi{Q}_0)$ are local, satisfy
$L_{\bi{Q}_j}(\mathfrak{R})=0$, and
commute, that is,
$[\bi{Q}_i,\bi{Q}_j]=0$, for all $i,j=0,1,2\dots$, and $\mathfrak{R}$
is hereditary on the linear span of $\bi{Q}_j$, $j=0,1,2,\dots$.
\looseness=-1
\end{theo}
\noindent{\it Proof.} The condition $L_{\bi{Q}_0}(\mathfrak{R})=0$
implies (cf.\
\cite{mik1,s}) that
\be\label{q0rep}
\bi{Q}_0'=\sum
_{j=2}^{q_0} \mathfrak{T}^{-1}
d_{j}^{(\bi{Q}_0)}
\tilde\mathfrak{R}^{j/r}\mathfrak{T}+\mathfrak{B}_{\bi{Q}_0},
\ee
where
$\mathfrak{B}_{\bi{Q}_0}\in
\mathrm{Mat}_{s}(\mathcal{A})[\![D^{-1}]\!]$
is a formal series with diagonal leading
coefficient, and $\deg\mathfrak{B}_{\bi{Q}_0}<2$;
$d_{j}^{(\bi{Q}_0)}$ are constant diagonal $s\times s$ matrices.
\looseness=-1 

From (\ref{q0rep}) it is immediate that $\nu=\mu q_0$.
It is also clear that 
$L_{\bi{S}}(\mathfrak{R})=\zeta\mathfrak{R}$, where
$\zeta=r\mu$. Using this formula 
and $[\bi{S},\bi{Q}_0]=\nu\bi{Q}_0=\mu
q_0
\bi{Q}_0$ 
yields $[\bi{S},\bi{Q}_j]=(\nu+j\eta)\bi{Q}_j=\mu (q_0+rj)
\bi{Q}_j$, whence
$L_{\bi{S}}(L_{\bi{Q}_j}(\mathfrak{R}))=L_{\bi{Q}_j}
(L_{\bi{S}}(\mathfrak{R}))+
\mu(q_0+r j)
L_{\bi{Q}_j}(\mathfrak{R})=\mu(r(j+1)+q_0)L_{\bi{Q}_j}(\mathfrak{R})$.

Using the Leibniz rule for the Lie derivative and
$L_{\bi{Q}_0}(\mathfrak{R})=0$, we obtain
$[\bi{Q}_0,\bi{Q}_j]=L_{\bi{Q}_0}(\bi{Q}_j)=L_{\bi{Q}_{0}}
(\mathfrak{R}^j\bi{Q}_{0})=L_{\bi{Q}_{0}}(\mathfrak{R}^j)
\bi{Q}_{0}+\mathfrak{R}^j[\bi{Q}_0,\bi{Q}_{0}]=
L_{\bi{Q}_{0}}(\mathfrak{R}^j)\bi{Q}_{0}=0$, whence
$L_{\bi{Q}_0}(L_{\bi{Q}_j}(\mathfrak{R}))=
L_{\bi{Q}_j}(L_{\bi{Q}_0}(\mathfrak{R}))=0$.
\looseness=-1 

Therefore, $L_{\bi{Q}_j}(\mathfrak{R})$ is a
formal symmetry of infinite rank for the system
$\bi{u}_{\tau}=\bi{Q}_0$. 
Now we can proceed inductively. 
Assume that we have already proved that $\bi{Q}_{j}$ is local 
(for $j=1$ this readily follows from our assumptions).
Then the coefficients of the formal series 
$L_{\bi{Q}_j}(\mathfrak{R})$ are local too, and hence 
$L_{\bi{Q}_j}(\mathfrak{R})=\sum
_{l=-\infty}^{k} \mathfrak{T}^{-1}
h_{l}^{(\bi{Q}_j)}\tilde\mathfrak{R}^{l/r}\mathfrak{T}$, where
$h_{l}^{(\bi{Q}_j)}$ are constant diagonal $s\times s$ matrices
and $k<r(j+1)+q_0$, cf.\ \cite{mik1,sok}. 

Using this formula and our
assumptions on $\bi{S},\bi{Q}_0$, and $\mathfrak{R}$, we readily
find that $L_{\bi{S}}(L_{\bi{Q}_j}(\mathfrak{R}))= \mu k
L_{\bi{Q}_0}(\mathfrak{R})$. But, as $k<r(j+1)+q_0$, this contradicts
our earlier result
$L_{\bi{S}}(L_{\bi{Q}_0}(\mathfrak{R}))=\mu(q_0+(j+1)r)
L_{\bi{Q}_j}(\mathfrak{R})$,
unless $L_{\bi{Q}_j}(\mathfrak{R})=0$. 
Hence, 
$L_{\bi{Q}_j}(\mathfrak{R})=0$.
Then, by the same argument as in the proof of Theorem~\ref{th_wh},
$\bi{Q}_{j+1}$ is local, and we can repeat the above reasoning
with $j$ replaced by $j+1$.

Thus, we have
$L_{\bi{Q}_j}(\mathfrak{R})=0$ for all $j\in\mathbb{N}$, and therefore
$\mathfrak{R}$ is hereditary on the linear span of
 $\bi{Q}_j=\mathfrak{R}^{j}(\bi{Q}_0)$, $j=0,1,2\dots$, so
$\bi{Q}_j$ commute for all $j=0,1,2,\dots$. 
$\square$

Note that the conditions of Theorem~\ref{th_sc}
are easy to verify in spite of their complicated appearance.
In particular, in the most common situation when
$\bi{S}=x\bi{u}_1+\Lambda\bi{u}$, where
$\Lambda$ is a constant diagonal matrix, we usually can choose
the coefficients of $\mathfrak{T}$ so that 
$L_{\bi{S}}(\mathfrak{T})=0$, and then in order to verify 
the equality
\be\label{scal}
L_{\bi{S}}(\mathfrak{T}^{-1}c\tilde\mathfrak{R}^{1/r}\mathfrak{T})=
\mu\mathfrak{T}^{-1}c\tilde\mathfrak{R}^{1/r}\mathfrak{T}
\ee 
for any
constant diagonal matrix $c$ it suffices to check that
$L_{\bi{S}}(\mathfrak{R})=
\zeta\mathfrak{R}$ for some nonzero constant $\zeta$.
Moreover, for $s=1$ we have $\mathfrak{T}=1$, and 
it suffices to check that $L_{\bi{S}}(\mathfrak{R})=
\zeta\mathfrak{R}$ for some $\zeta=\mathrm{const}\neq 0$
in order to prove (\ref{scal}) even if
$\bi{S}$  is not of the form $x\bi{u}_1+\Lambda\bi{u}$. 

Consider now an application of our results to operators other
than recursion operators. Recall that
$\mathfrak{K}\in\mathrm{Mat}_{s}(\mathcal{A})[\![D^{-1}]\!]$ is
called \cite{ff} an {\em inverse Noether operator} for an
evolution system $\bi{u}_{t}=\bi{F}(x,t,\bi{u},\dots,\bi{u}_n)$
(respectively $\bi{u}_{\tau}=\bi{F}(x,t,\bi{u},\dots,\bi{u}_n)$,
where $\tau$ is an evolution parameter other than $t$), if it
satisfies the equation $\p\mathfrak{K}/\p t +
\mathfrak{K}'[\bi{F}] +(\bi{F}')^{\dagger}\circ
\mathfrak{K}+\mathfrak{K}\circ \bi{F}'=0$ (resp.\
$\mathfrak{K}'[\bi{F}] +(\bi{F}')^{\dagger}\circ
\mathfrak{K}+\mathfrak{K}\circ \bi{F}'=0$).
%
%
We shall restrict ourselves to considering the inverse Noether
operators of the form \be\label{ino}
\mathfrak{K}=\sum\limits_{i=0}^{h}b_{i}D^{i}+\sum
\limits_{\alpha=1}^{l}\boldsymbol{\eta}_{\alpha}\otimes
D^{-1}\circ \boldsymbol{\zeta}_{\alpha}, \ee where $b_i$ are
$s\times s$ matrices with entries from $\mathcal{A}$, and
$\boldsymbol{\eta}_{\alpha},
\boldsymbol{\zeta}_{\alpha}\in\mathcal{A}^{s}$.
%
The majority of known today inverse Noether operators of
integrable (1+1)-dimensional evolution systems indeed are of the form
(\ref{ino}), see e.g.~\cite{wang}.
\looseness=-2

Our Theorems~\ref{th_wh} and \ref{th_sc} readily imply the following
assertion. 
\begin{cor}\label{cor_wh}
Let 
a system $\bi{u}_t=\bi{F}(x,t,\bi{u},\dots,\bi{u}_n)$ or
$\bi{u}_{\tau}=\bi{F}(x,t,\allowbreak\bi{u},
\dots,\allowbreak\bi{u}_n)$,
$\tau\neq t$,  have a recursion operator $\mathfrak{R}$ 
meeting the requirements of Theorem~\ref{th_wh} or \ref{th_sc} and
an inverse Noether operator $\mathfrak{K}$ of 
the form (\ref{ino}) with
$h\geq 0$. Assume that there exist the functions 
$d_{\alpha,i,a}(t)$ such
that $\boldsymbol{\zeta}_{\alpha}= \sum
_{i=-1}^{\infty}\sum
_{a=1}^{s}d_{\alpha,i,a}(t)\delta\rho_{i}^{a}/\delta\bu$ for all
$\alpha=1,\dots,l$, where $\rho_i^a$ are canonical densities
associated with
$\mathfrak{R}$.

Then
$\boldsymbol{\chi}_{j}\equiv\mathfrak{K}(\bi{Q}_j)$ are local
(i.e., $\boldsymbol{\chi}_{j}\in\mathcal{A}^{s}$) for all
$j=0,1,2,\dots$.
\end{cor}
\noindent{\it Proof.} We have
$\boldsymbol{\zeta}_{\alpha}\cdot\bi{Q}_j\in\Im D$  by (\ref{gammas}),
because
$\delta\rho_{i}^a/\delta\bu\cdot\bi{Q}_j\in\Im D$, cf.\ the proof of
Theorem~\ref{th_wh}. Therefore, 
$D^{-1} (\boldsymbol{\zeta}_{\alpha}\cdot{\bi{Q}}_j)\in\mathcal{A}$ for
all
$\alpha=1,\dots,p$ and $j=0,1,2\dots$, and thus
$\mathfrak{K}(\bi{Q}_j)\in\mathcal{A}^{s}$.
$\square$
\looseness=-1
\section{Locality of hierarchies of symmetries\\ generated by
nonhereditary recursion operators}
%
Even if a recursion operator
is not hereditary, it still
can generate a hierarchy of local symmetries.
%
A simple example is given  (see \cite{nhro}
for a more complicated example with $s=3$) by
\be\label{nh0}
\mathfrak{R}=\left(\begin{array}{cc} D & u_1 D^{-1}\\ 0 & -D+v
\end{array}
\right). \ee
%
For $\bi{K}=(u,0)^{T}$ we have
$L_{\bi{K}}(\mathfrak{R})=0$ but
$L_{\mathfrak{R}(\bi{K})}(\mathfrak{R})\neq 0$, so
$\mathfrak{R}$ (\ref{nh0}) is not hereditary. Nevertheless, all
$\bi{K}_i=\mathfrak{R}^i(\bi{K})$
obviously are local:
assuming that $D^{-1}(0)=0$, see e.g.\ \cite{gut, sw1, serg} for
more details on defining $D^{-1}$, we readily find that
$\bi{K}_i=(u_i,0)^{T}$.
\looseness=-1

While in the above example the locality of
symmetries was obvious, 
this is not the case in general.
Fortunately, there is a way around: in Theorem~\ref{th_loc} below we
present the sufficient conditions ensuring the locality of
$\bi{Q}_i=\mathfrak{R}^{i}(\bi{Q}_0)$ without assuming that
$\mathfrak{R}$ is hereditary.
To this end we need, however, to fix an evolution system
\be\label{eveq}
\p\bu/\p t=\F(x,t,\bu,\bu_1,\dots,\bu_n),\quad n\geq 2,
\ee
with
$\p\F/\p\bu_n\neq 0$.

If the matrix $\p\F/\p\bu_n$ is nondegenerate and has $s$
distinct eigenvalues, we shall say that (\ref{eveq}) is a {\em
nondegenerate weakly diagonalizable (NWD)} system.
The properties of such systems were studied e.g.\ in
\cite{mik1,s,srni}. Clearly,
for $s=1$ any system (\ref{eveq}) with $\p\F/\p\bu_n\neq 0$ is NWD.
\looseness=-1

Recall that $\bi{G}\in\mathcal{A}^{s}$ is called \cite{olv_eng2,
s} a {\em symmetry} for (\ref{eveq}) if
$\p\bi{G}/\p t+[\F,\G]=0$,
where $[\cdot, \cdot]$ is the Lie bracket, defined above.
$$
\hspace*{-54mm}\mbox{Let}\,\,n_0=\left\{\begin{array}{l} 1-j,\,
\mbox{where
$j$
is the greatest number such that}\\
\,\,\p\F/\p\bu_i\,\,\mbox{for}\,\,
i=n-j,\dots,n\,\,\mbox{depend
on
$x$ and $t$ only},\\ 2\,\,\mbox{otherwise.} \end{array} \right.
$$
%
%
Now assume that (\ref{eveq}) is NWD,
$\mathfrak{R}$ is the recursion operator for (\ref{eveq}),
and $\bi{Q}_0\in\mathcal{A}^s$ is a 
symmetry for (\ref{eveq}),
and hence
all $\bi{Q}_i=\mathfrak{R}^{i}(\bi{Q}_0)$ are symmetries for
(\ref{eveq}). Further assume that $\p\bi{Q}_0/\p t=0$ and that
$\mathfrak{R}$ is such that 
this implies
$\p\bi{Q}_j/\p t=0$ for all $j\in\mathbb{N}$.

Further assume that (\ref{eveq}) has a time-independent
($\p\mathfrak{L}/\p t=0$) nondegenerate
formal symmetry $\mathfrak{L}=\sum
_{j=-\infty}^{m}g_{i}D^i$ of infinite rank
and of
degree $m>0$, and the matrix $g_m$ has
$s$ distinct eigenvalues $\lambda_i$.
Clearly, if $r>0$, $\p\mathfrak{R}/\p t=0$, 
$\det a_r\neq 0$ and $a_r$ has
$s$ distinct eigenvalues,  we can 
take $\mathfrak{R}$ for $\mathfrak{L}$.
\looseness=-1

Then, 
using the
diagonalizing transformation $\mathfrak{T}$ for
$\mathfrak{H}=\mathfrak{L}$ from Lemma~\ref{lem1},
we readily see that if $\bi{Q}_j$
is local, then $\bi{Q}'_j$ is a
formal symmetry of degree $q_j\equiv\ord\bi{Q}_j$ 
and of rank $q_j-n_0+2$
for (\ref{eveq}), cf.\ e.g.\ \cite{mik1, srni}. 
Hence we have
$\bi{Q}_j'=\sum
_{k=n_0}^{q_j} \mathfrak{T}^{-1}
d_{k}^{(\bi{Q}_j)}\tilde\mathfrak{L}^{j/m}\mathfrak{T}
+\mathfrak{B}_{\bi{Q}_j}$,
where $\deg\mathfrak{B}_{\bi{Q}_j}<n_0$,
$\tilde\mathfrak{L}=\mathfrak{T}\mathfrak{L}\mathfrak{T}^{-1}$ and
$d_{k}^{(\bi{Q}_j)}$ are constant diagonal $s\times s$ matrices,
cf. e.g.~\cite{mik1, s, srni}. 
Using this result, we can show that $\deg
L_{\bi{Q}_j}(\mathfrak{L})\leq m-1+2-n-n_0$, hence $\mathfrak{L}$
is a formal symmetry of rank $q_j+n-n_0+1$ for the system
$\bu_{t_j}=\bi{Q}_j$, where $q_j=\ord\bi{Q}_j$, $j=0,1,2\dots$.
Indeed, for $s=1$ this holds by Lemma 11 of Sokolov \cite{sok}, 
and for $s>1$ this is proved
along the same lines as in \cite{sok}.

Hence, by the results of \cite{mik1}, for $a=1,\dots,s$ and
$b=-1,0,1,\dots, n-n_0-1$ the canonical densities $\rho_{b}^{a}$
associated with
$\mathfrak{L}$
are
conserved densities for the systems
$\p\bu/\p t_j=\bi{Q}_j$,
$j=0,1,2,\dots$, i.e., ${\rho'}^a_b[\bi{Q}_j]\in\Im D$, and
the locality of all $\bi{Q}_k$, $k\in\mathbb{N}$, is proved
by induction on $k$ as in the proof of Theorem~\ref{th_wh}, 
so we arrive
at the following result. \looseness=-2
\begin{theo}\label{th_loc}
Let an NWD system (\ref{eveq}) with $n\geq 2$
have a time-in\-de\-pen\-dent nondegenerate formal symmetry
$\mathfrak{L}$ of degree $m>0$ and of infinite rank, and let the
leading coefficient of $\mathfrak{L}$ have $s$ distinct
eigenvalues. Further assume that (\ref{eveq}) has a time-independent
symmetry $\bi{Q}_0\in\mathcal{A}^s$,
$\ord\bi{Q}_0\geq\max(n_0,0)$, and a recursion operator
$\mathfrak{R}$ of the form (\ref{ro}) with $r\geq 0$ such that 
$\bi{Q}_i=\mathfrak{R}^i(\bi{Q}_0)$ are time-independent for all
$i\in\mathbb{N}$.
Suppose that there exist the functions $c_{\alpha,i,a}(t)$
such that $$\bg_{\alpha}= \sum
_{i=-1}^{n-n_0-1}\sum
_{a=1}^{s}c_{\alpha,i,a}(t)\delta\rho_{i}^{a}/\delta\bu$$ for all
$\alpha=1,\dots,p$, where $\rho_i^a$ are canonical
densities associated with $\mathfrak{L}$.
\looseness=-2

Then $\bi{Q}_{j}$ are local, that is,  
$\bi{Q}_{j}\in\mathcal{A}^{s}$, 
for all $j\in\mathbb{N}$.
\end{theo}

Let us mention that, exactly as in the case 
of one dependent variable $u$, 
if $\p\F/\p t=0$, then the densities
$\rho_i^a$ for
$i=-1,\dots,n-n_0-1$
can be computed directly from $\F$. 
Namely, up to multiplication by a constant and adding a linear
combination of $\rho_k^a$, $k<j$, we have 
$\rho_{0}^{a}=\res\ln
((D_{t}(\mathfrak{T})\mathfrak{T}^{-1}
+\mathfrak{T}\F'\mathfrak{T}^{-1})_{aa})^{1/n}$, and
$\rho_{j}^{a}=
\res((D_{t}(\mathfrak{T})\mathfrak{T}^{-1}+
\mathfrak{T}\F'\mathfrak{T}^{-1})_{aa})^{j/n}$,
$j=-1,1,2,\dots, n-n_0-1$, 
$a=1,\dots,s$.
\looseness=-1

In complete analogy with Corollary~\ref{cor_wh}, 
we have the following result.
\begin{cor}\label{cor_loc}
Under the assumptions of Theorem~\ref{th_loc}, let 
(\ref{eveq}) have an inverse Noether operator $\mathfrak{K}$ of
the form (\ref{ino}) with $h\geq 0$. Assume that there exist the
functions $d_{\alpha,i,a}(t)$ such that
$$\boldsymbol{\zeta}_{\alpha}= \sum
_{i=-1}^{n-n_0-1}\sum
_{a=1}^{s}d_{\alpha,i,a}(t)\delta\rho_{i}^{a}/\delta\bu$$ for all
$\alpha=1,\dots,l$, where $\rho_i^a$ are canonical densities
associated with $\mathfrak{L}$.

Then
$\boldsymbol{\eta}_{j}=\mathfrak{K}(\bi{Q}_j)$ are local,
i.e., $\boldsymbol{\eta}_{j}\in\mathcal{A}^{s}$, for all
$j=0,1,2,\dots$.
\end{cor}


\section{Examples}
Let us agree that for scalar
$\bi{u}$, i.e., for $s=1$,
we shall write $u,u_j,\gamma_\alpha,F,Q$ instead of
$\bu,\bu_j,\boldsymbol{\gamma}_\alpha,\F,\bi{Q}$, as we already did in
Introduction.
%

Consider first the generalized Korteweg--de Vries equation
\cite{chou} \be\label{gkdv} u_t=-u_3-6 u u_1-6 f(t) u+ x (\dot
f(t)+ 12(f(t))^2), \ee where $f(t)$ is an arbitrary smooth
function of time $t$.

This equation has \cite{chou} a hereditary recursion operator
$$
\mathfrak{R}=(g(t))^2 (D^2+4 (u-xf(t)))+2 g(t)
(u_1-f(t))D^{-1}\circ g(t),
$$
where
$g(t)=\exp(6\int_{t_0}^{t} f(t')dt')$. Clearly, $\mathfrak{R}$ is
of the form (\ref{ro}) with $p=1$ and $\gamma_1=g(t)$. Upon setting
$\mathfrak{H}=\mathfrak{R}$ we find that
$\rho_1=2 g(t)(u-xf(t))$, so $\delta\rho_1/\delta u=2 g(t)$.
Taking $c_{1,1}(t)=1/2$, we see that the requirements of
Theorem~\ref{th_wh_sc} are met, and hence the repeated application of
$\mathfrak{R}$ to the seed symmetry $Q=g(t) (u_1-f(t))$ yields 
an infinite
hierarchy of local symmetries for (\ref{gkdv}). \looseness=-1

Let now $s=2$, $\bi{u}=(u,v)^{T}$,
and consider the
so-called  DS IV system \cite{ds_art} 
(also known as the Hirota--Satsuma \cite{hs}
system)
\be\label{ds}
\ba{l}
u_t=u_3/2 + 3 u u_1- 6  v v_1,\\
v_t=-v_3-3 u v_1.
\ea
\ee 
and its recursion operator, see e.g.\ \cite{sok99},
\be\label{dsro}
\ba{l}
\mathfrak{R}=\left(\begin{array}{cc}  \mathfrak{B} & - 5v D^2-4 v_1
D-v_2 - 4 u v\\ -(5/2)v_1 D- 3 v_2 & \mathfrak{C}
\end{array}
\right)\\[4mm]
+\G_1 \otimes D^{-1}\circ\bg_1+\G_2 \otimes D^{-1}\circ\bg_2,
\ea
\ee
where $\mathfrak{B}=D^4/2+2u D^2+3 u_1 D+
2 u_2+4(u^2-v^2)$, $\mathfrak{C}= -D^4- 4 u D^2- 2
u_1 D -4 v^2$, $\G_1=(u_1, v_1)^{T}$, $\bg_1=(u, -2 v)$,
$\G_2=(u_3/2 + 3 u u_1- 6  v v_1,
-v_3-3 u v_1)^{T}$, $\bg_2=(1, 0)$.

Set $\bi{Q}_0=(u_3/2 + 3 u u_1- 6  v v_1, -v_3-3 u v_1)^T$
and $\bi{S}=x\bi{u}_1+ 2\bi{u}$.
Then all requirements
of Theorem~\ref{th_sc} (and of Theorem~\ref{th_wh} too, as 
$\mathfrak{R}$ is
hereditary) are met. In particular, there exist
the constants $c_{1,1,1}$,
$c_{1,3,1}$, $c_{1,3,2}$ such that
$\bg_1=c_{1,1,1}\delta\rho^1_1/\delta\bi{u}
+c_{1,3,1}\delta\rho^1_3/\delta\bi{u}
+c_{1,3,2}\delta\rho^2_3/\delta\bi{u}$ and
$\bg_2=c_{2,1,1}\delta\rho^1_1/\delta\bi{u}$. 
Thus, by Theorem~\ref{th_sc} all
$\bi{Q}_j=\mathfrak{R}^j(\bi{Q}_0)$ are local.

Moreover, the system (\ref{ds}) has an inverse Noether operator
which also is symplectic, cf.\ e.g.\ \cite{ds_art,fu82},
$$
\mathfrak{K}=\left(\begin{array}{cc} D/2 & 0\\ 
0 & -2 D 
\end{array}
\right)
+\bg_1 \otimes D^{-1}\circ\bg_2
+\bg_2 \otimes D^{-1}\circ\bg_1.
$$
It is immediate that the requirements of
Corollary~\ref{cor_wh} are met, and thus the cosymmetries
$\boldsymbol{\chi}_{j}=\mathfrak{K}(\bi{Q}_j)$ are
local for all $j=0,1,2,\dots$. 

Consider now the Calogero--Degasperis--Fokas equation
\cite{cd, fokas} \be\label{cdf} u_t=u_3-u_1^3/8+(a\exp(u)
+b\exp(-u)+c)u_1, \ee where $a,b,c$ are arbitrary constants. It
has the recursion operator \cite{fokas} 
$$
\ba{l}
\mathfrak{R}=D^2-{\ds\frac{u_1^2}{4}}+{\ds\frac{2}{3}} (a\exp(u)
+b\exp(-u)+c)\\[3mm]
 +{\ds\frac{u_1}{3}} D^{-1}\circ \left({\ds\frac{3
u_2}{4}}+a\exp(u) -b\exp(-u)\right). 
\ea
$$ 
For
$\mathfrak{H}=\mathfrak{R}$ we have $\rho_{1}^{1}=-u_1^2/8+(1/3)
(a\exp(u) +b\exp(-u)+c)$, so $\gamma_1=3\delta\rho_1^1/\delta u$.
Thus, $\mathfrak{R}$ meets the requirements of
Theorem~\ref{th_loc}  (and of Theorem~\ref{th_wh} as well,
because $\mathfrak{R}$ is hereditary), and hence all
$Q_j=\mathfrak{R}^j (u_1)$, $j\in\mathbb{N}$, are local.

Let again $s=2$, $\bi{u}=(u,v)^T$. Consider the operator
\cite{bl_art}
\be\label{bl_ro}
\mathfrak{R}=\left(\begin{array}{cc} D^2/4+v-u^2/4 & 3 D/2+ u/2
\\  (3/4)v_1 & D^2 + u D+v
\end{array}
\right)+\sum\limits_{\alpha=1}^{2}\G_\alpha \otimes
D^{-1}\circ\bg_\alpha,
\ee
where $\G_1=(2 v_1, v_2+ u v_1)^{T}$,
$\bg_1=(1/2, 0)$,
$\G_1=(u_1, v_1)^{T}$, $\bg_2=(-u/4, 1/2)$.
This is a recursion operator for the NWD system
\be\label{bl_sys}
\ba{l}
u_t=u_3/4+ (3/2)(v_2+ v u_1+ u v_1)-(3/8)u^2 u_1,\\ 
v_t=v_3+ (3/2) (u v_2 +v v_1) +(3/8) u^2 v_1.
\ea
\ee

We have $\rho_0^2=u$ and
$\rho_1^1=v/2-u^2/8$, so
$\bg_1=(1/2)\delta\rho_0^1/\delta\bi{u}$ and
$\bg_2=\delta\rho_1^1/\delta\bi{u}$.  
Hence, $\mathfrak{R}$ meets the requirements of Theorem~\ref{th_loc}
for $\F=\bi{Q}_0=\mathfrak{R}(\bi{u}_1)$, where $\bi{F}$ stands for the
right-hand side of (\ref{bl_sys}), and thus all
$\bi{Q}_j=\mathfrak{R}^{j+1}(\bi{u}_1)$, $j\in\mathbb{N}$, are local.
\looseness=-1

%

Let us stress that
(\ref{gkdv}), (\ref{cdf})  and (\ref{bl_sys}) have
no scaling symmetry of the form used in \cite{swro}, 
so in these cases it
is impossible prove the locality of hierarchies of symmetries 
using the results of
\cite{swro}. 
  
\subsection*{Acknowledgements}
I am sincerely grateful to Profs. B. Fuchssteiner and V.V. Sokolov
for stimulating discussions.
I am also pleased to thank Dr. M. Marvan 
for kindly reading the
manuscript of the present paper and making a number of useful remarks. 
I acknowledge with gratitude the kind hospitality of
organizers of Banach Center Workshop ``Multi-Hamiltonian
Structures: Algebraic and Geometric Aspects'' in B\c{e}dlewo,
Poland, where some of the results of this paper were presented.
\looseness=-1

This research was supported by DFG via Graduiertenkolleg
``Geometrie und Nichtlineare Analysis'' of Institute f\"ur
Mathematik of Hum\-boldt-Universit\"at zu Berlin, Germany, where the
author held  a postdoctoral fellowship. I also acknowledge the
partial support from the Ministry of Education, Youth and Sports
of Czech Republic under Grant MSM:J10/98:192400002,  and from the
Czech Grant Agency under Grant No. 201/00/0724.
\looseness=-1


\end{document}